\documentclass[aps,twocolumn,superscriptaddress,showpacs]{revtex4-1}

\usepackage{natbib}  
\usepackage{amssymb}

\usepackage{comment}
\usepackage{graphicx}

\begin{document}

\title{A Spallation Model for the Titanium-rich Supernova Remnant Cassiopeia A}
\date{\today}
\author{Rachid Ouyed}
\affiliation{Department of Physics and Astronomy, University of Calgary, 
2500 University Drive NW, Calgary, Alberta T2N 1N4 Canada}
\author{Denis Leahy}
\affiliation{Department of Physics and Astronomy, University of Calgary, 
2500 University Drive NW, Calgary, Alberta T2N 1N4 Canada}
\author{Amir Ouyed}
\affiliation{Department of Physics and Astronomy, University of Calgary, 
2500 University Drive NW, Calgary, Alberta T2N 1N4 Canada}
\author{Prashanth Jaikumar}
\affiliation{Department of Physics and Astronomy, California State University  Long Beach,\\ 1250 Bellflower Blvd., Long Beach, CA 90840 USA}


\begin{abstract} 
Titanium-rich subluminous supernovae are rare and challenge current SN nucleosynthesis models. We present a model in which ejecta from a standard Supernova  is impacted by a second explosion of the neutron star (a Quark-nova), resulting in  spallation reactions that lead to $^{56}$Ni destruction and $^{44}$Ti creation under the right conditions. Basic calculations of the spallation products shows
that a delay between the two explosions of $\sim$ 5 days reproduces the observed abundance of $^{44}$Ti in Cas A and explains its low luminosity as a result of the destruction of $^{56}$Ni. Our results could have important implications
for lightcurves of subluminous as well as superluminous supernovae.
 \end{abstract}
\maketitle

\section {Introduction} Cas A is a young galactic supernova remnant formed in the aftermath of a core-collapse explosion of a massive star. Tentative classification as TypeIIb (based on the detection of weak Helium lines in addition to the Hydrogen line) has been strengthened by infra-red studies of the scattered light echo \citep{krause08}. Cas A is intensively studied as a prototype that can unveil the pristine composition of nucleosynthetic yields, thereby constraining aspects of the supernova mechanism and stellar evolution models. Multi-wavelength studies show that emission from the remnant is dominated by a bright ring, but also find jets and knots, as well as an X-ray bright CCO (central compact object) believed to be a rapidly cooling neutron star. Cas A is an unusual supernova (SN) in some respects: nuclear decay lines of $^{44}$Ca (1157 keV) and $^{44}$Sc (67.9 and 78.4 keV) detected by COMPTEL and BEPPO SAX indicate a very large synthesized $^{44}$Ti mass of (0.8-2.5)$\times 10^{-4}M_{\odot}$ \citep{iyudin94} ($M_{\odot}$ is the solar mass), which would imply an ejected $^{56}$Ni mass of at least 0.05 $M_{\odot}$ \citep{woosley95,magkotsios10}, making Cas A extremely bright given its proximity; but no definitive historical record of such a bright SN at that time ($\sim$ 300 yrs ago) exists. The possible detection by Flamsteed in 1680 suggests 6th magnitude. Possible solutions center on extinction due to a surrounding dust cloud, possibly generated by a pre-SN wind from the massive star and an asymmetric SN explosion \citep{maeda03}. Although observations of Cas A do give indications of these features, they do not explain why no other
Ti-rich SNe have been found despite searches in massive star regions in the inner galaxy \citep{the06,diehl06}. It appears that Ti-producing SNe are quite rare. NuSTAR (nuclear spectroscopic telescope array) aims to map the $^{44}$Ti in more Cas A-like remnants to solve this puzzle.

In this letter, we present an alternative that can reconcile
the sub-luminous nature of Cas A with its excess $^{44}$Ti production. Assuming that the progenitor of the Cas A remnant was not atypical (i.e., did not produce
large amount of Ti in situ), we suggest that Ti is formed instead as a
spallation product when neutron-rich material ejected from a "second" explosion , viz., that of the neutron star (a quark-nova), impacts and destroys the $^{56}$Ni layer ejected by the preceding explosion (the SN).  The basic picture  is that a SN can produce a massive neutron star, which then converts 
explosively to a quark star (QS hereafter) (in an event called a Quark Nova or QN; \citep{ODD,KOJ}).
Such an explosion can happen if the Neutron Star (NS), in its spin-down evolution, reaches the quark deconfinement density and subsequently undergoes a phase transition to the more stable strange quark matter phase
 \citep{itoh70,bodmer71,witten84}, resulting in a conversion front that propagates toward the surface in the detonative regime.
The outer layers of the parent  NS are ejected  from an expanding thermal fireball \citep{vogt04,ouyed05} which  allows for ejecta with kinetic energy easily exceeding $10^{52}$ erg.
If the QN occurs less than a few weeks after the SN, the QN energy release reheats the preceding SN ejecta,
and results in an event we call a dual-shock Quark Nova (dsQN hereafter).
In previous papers, we introduced the dsQN as a model for superluminous SNe \cite{leahy08} and discussed 
 their photometric  and spectroscopic \citep{ouyed10} signatures. 
 Here, for the first time, we explore the nuclear  processing of the innermost SN ejecta by the QN relativistic ejecta (neutrons and heavy nuclei).  We show that such a model for Cas A can explain why the SN is sub-luminous yet Ti-rich; in addition, the rarity of such events follows naturally as a constraint from the time delay between the two explosions. Testable predictions   based on our model are, delayed Hydrogen signatures (weeks after the second explosion) and a modified SN light curve.

\section{The spallation model}
{\it Beam and Target}: In analogy with spallation reactions in the laboratory, we frame our model in the context of a {\it "beam"} and a {\it "target"}. The QN provides the "beam": a relativistic outflow of neutron-rich material from the NS surface, caused by an explosive  phase transition in its core to a more compact quark phase. Recent numerical studies of the phase conversion front \citep{niebergal10} suggest supersonic laminar motion of the conversion front, which can become unstable \citep{Hor10}, wrinkling the conversion front to serve as a platform for a DDT (deflagaration-to-detonation). The outcome, depending on the conversion efficiency of the shock to kinetic energy, is ejection of about $M_{\rm QN}\sim 10^{-3}M_{\odot}$ \citep{ODD,KOJ} of the NS's outermost layers at nearly relativistic speeds with average Lorentz factors of $\Gamma_{\rm QN}\sim 10$. The total number of ejected nucleons in this beam (mostly neutrons) is then $N^{0}\sim 1.2\times 10^{54} M_{\rm QN, -3}$ where $M_{\rm QN, -3}$ is the QN ejecta mass in units of $10^{-3}M_{\odot}$. Adopting these fiducial values, the neutron energy is $E_0\sim 10$ GeV. This beam of relativistic neutrons (speeds close to $c$), will overtake and strike the innermost layers of expanding SN ejecta (the "target") which are moving much slower at a speed "$v\ll c$". Setting our clock by the SN explosion at $t$=0, this collision will happen a time $t_{\rm delay}$, the delay between the SN and the QN explosions. The collision between the QN and SN ejecta causes spallation and subsequently other nuclear reactions. The crux of our argument is that these spallation reactions can be a mechanism to explain some of the unique features of Cas A discussed previously, if $t_{\rm delay}$ is chosen appropriately.

We assume an onion-like profile of the expanding shocked SN ejecta (i.e., no mixing) with the innermost ejecta, viz., Ni nuclei (mass number $A$=$56$) constituting the target at a distance from the CCO of $R_{\rm  in}(t)$=$ v\,t_{\rm delay}$. The target number density in the Ni layer is approximately constant at $n_{\rm A}$=$M_{\rm A}/(4\pi R_{\rm in}^2 \Delta R)$, where $\Delta R$ is the thickness of the Ni layer. The neutron mean free path for spallation in the Ni layer is $\lambda=1/(n_{\rm A}\sigma_{\rm sp})$, where the spallation cross-section for neutrons on a target nucleus is empirically described with $\sigma_{\rm sp}\simeq 45 A^{0.7}\,f(A)$ mb with $f(A)\sim 1$ a factor of order unity \citep{Letaw83}. We have ignored a weak dependence on energy ($\le 10\%$ for energies $E>100$ MeV; \cite{Letaw83}), so that spallation mean free paths for the few subsequent generations in the spallation cascade are roughly constant. 
The average number of collisions an incoming neutron makes in the Ni layer is 
\begin{equation} N_{\rm coll.}\approx\frac{\Delta R}{\lambda}\simeq 2.75 \frac{ M_{A,0.1}}{{\left(A_{56}\right)}^{0.3}(v_{5000\,\rm{km/s}}\, t_{\rm delay, 5 \,{\rm days}})^2} \ ,
\label{Ncoll}
\end{equation}
where $M_{\rm A}$ is in units of $0.1M_{\odot}$.
The  SN ejecta must be sufficiently dense that $N_{\rm coll.} \geq 1$, which for fiducial values of $M_{A}$ and $v$ in Eq.(\ref{Ncoll}), limits $t_{\rm delay} < 8.3$ days.

{\it Spallation Reactions}: We include only the two most relevant reactions: spallation by neutrons and by protons (e.g. $n + A \rightarrow products + n\times (\zeta_{\rm nn}+1)+  p \times \zeta_{\rm np}$) where $\zeta$ is the multiplicity. The $n+A$ neutron multiplicity  as a function of the beam energy and target material shows roughly linear dependence on the target mass number (in the range $12 < A < 238$). We use the semi-empirical formula $\zeta_{\rm nn} (E)\simeq A\, (0.0833+0.0317 \ln{E})$, where the  neutron energy $E$ is in GeV\citep{Cugnon97}. This formula gives better than 10\% accuracy for $A > 40$. The average total multiplicity is 
\begin{equation} \label{eq:zetaprimary} \bar{\zeta} (E,A) =   \zeta_{\rm nn}+\zeta_{\rm np}   \simeq  4.67 A_{56}(1+ 0.38\ln{E})Y_{\rm np} \ , \end{equation}
where $Y_{\rm np} =(1+ \zeta_{\rm np}/\zeta_{\rm nn})$ is in the range  $1.25 < Y_{\rm np}  < 1.67$ (e.g.  \cite{Cugnon97}). For  proton induced reactions ($p+A$), most of the results are  similar to those of neutron induced reactions \citep{Cugnon97}.  We treat neutrons and protons identically. Spallation effectively ceases when $\bar{\zeta}$ drops below 1, corresponding to projectile energy of $E\sim 73$ MeV.

\begin{figure*}[t!] \centering \includegraphics[width=0.325\textwidth]{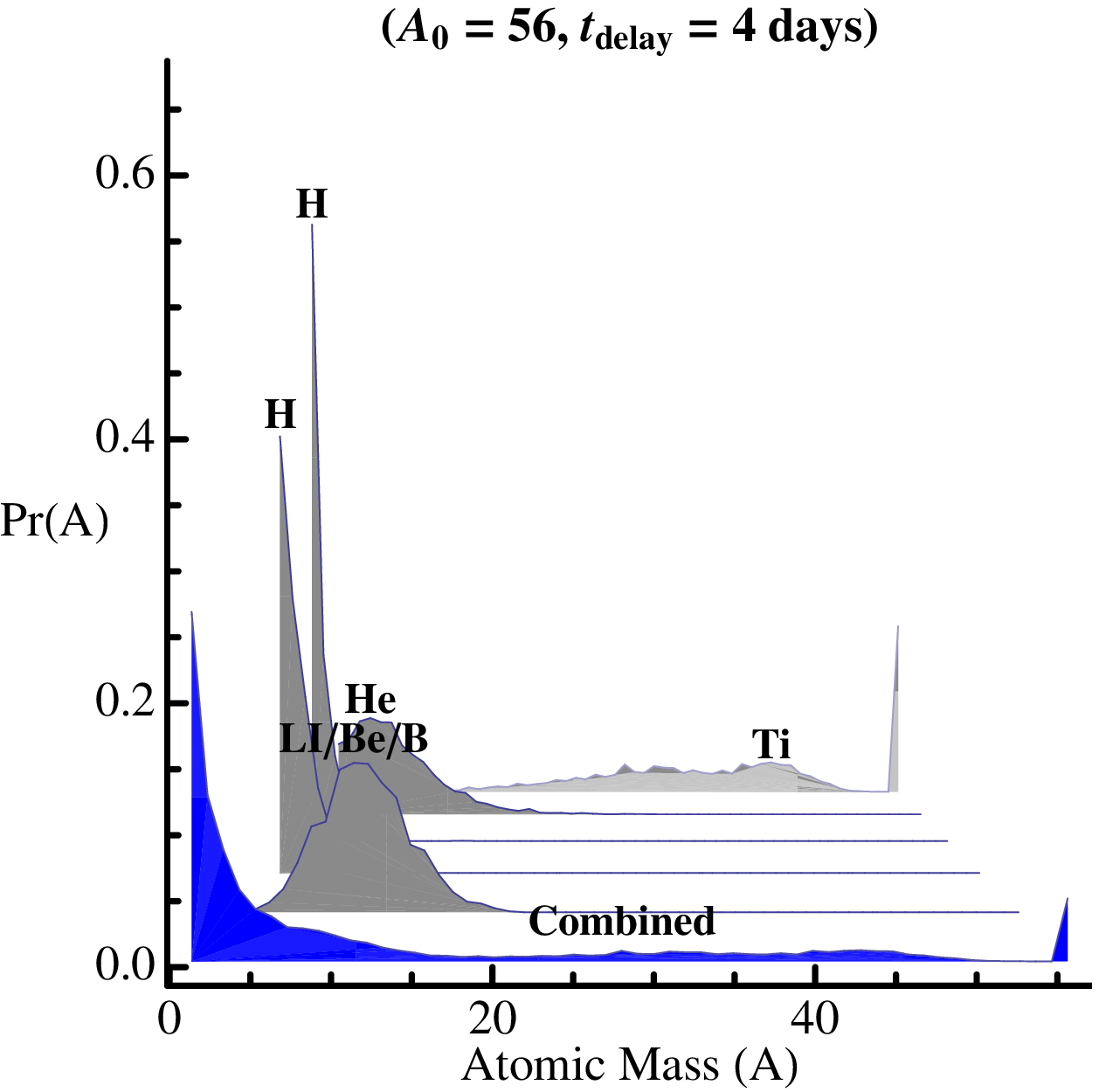} \includegraphics[width=0.325\textwidth]{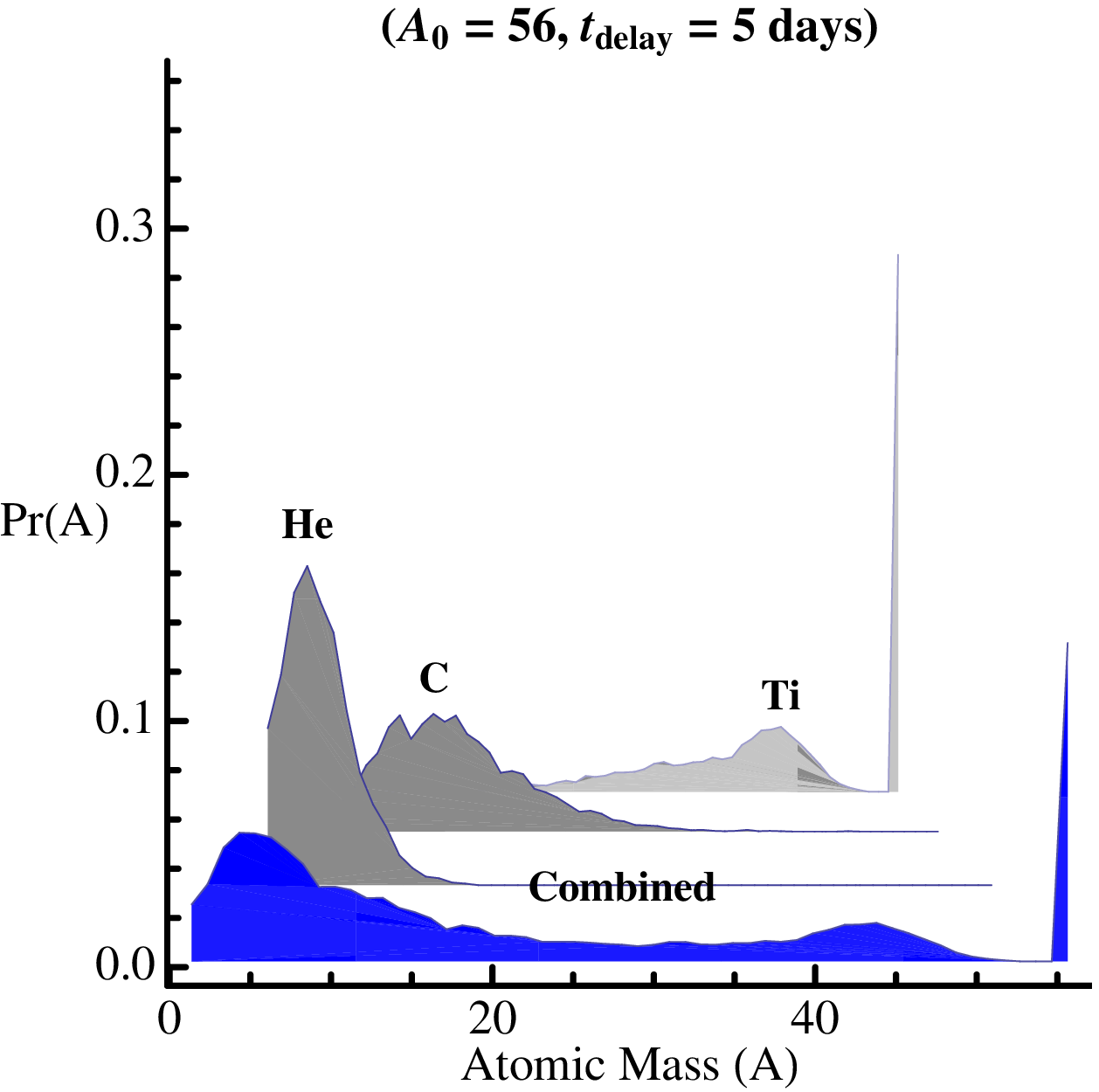} \includegraphics[width=0.325\textwidth]{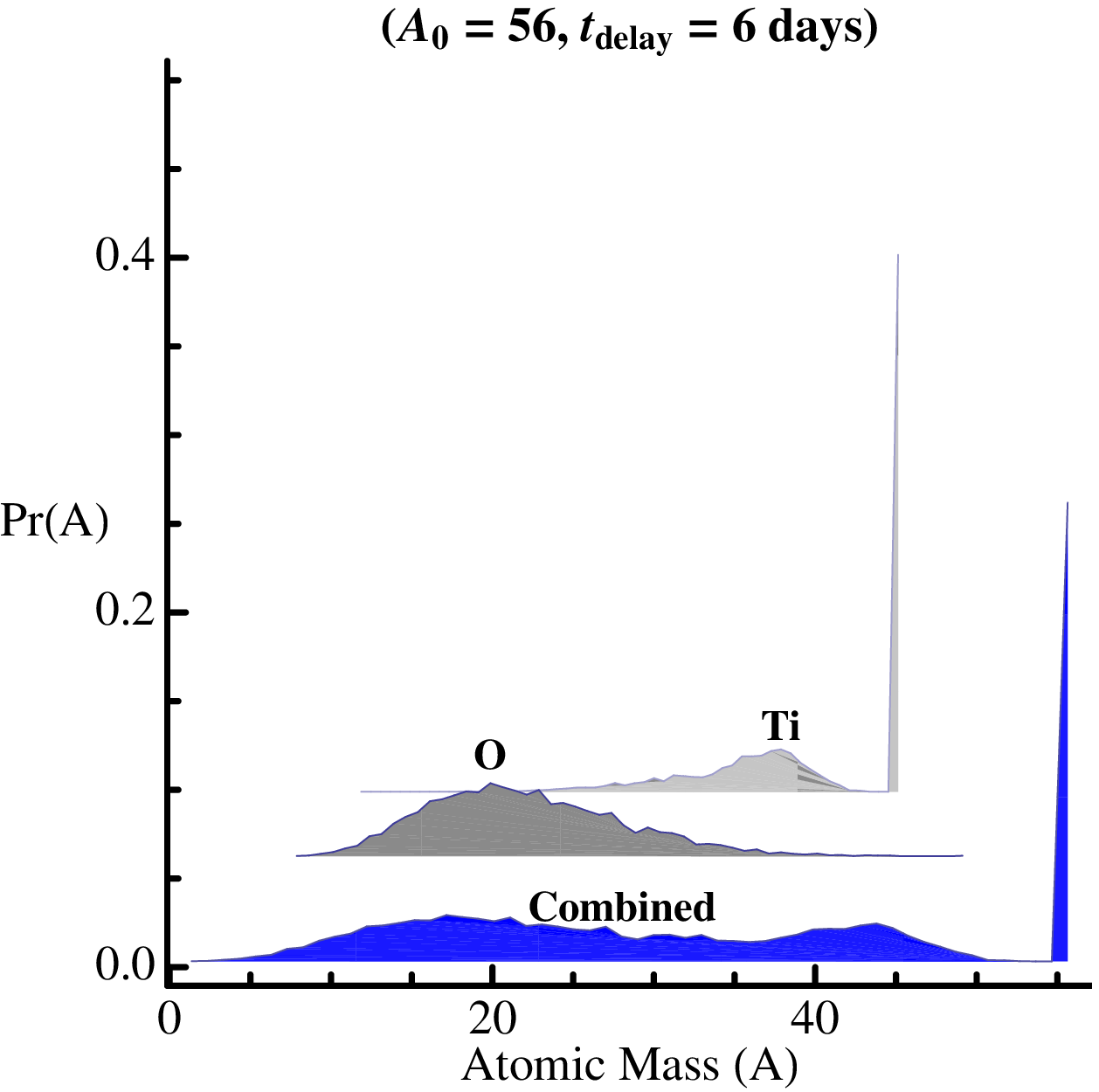} \caption{Spallation products in successive layers (back to front) from $^{56}$Ni  for $t_{\rm delay}=4,5,6$ days. The overall distribution is shown in the front layer labelled  ``combined".} \label{generations} \end{figure*}

{\it Spallation Statistics}: Let us divide the Ni layer into $\sim N_{\rm coll}$ imaginary sub-layers of radial thickness $\sim \lambda$, with $k$=0 denoting the innermost one. A given Ni nucleus in this layer will be hit multiple times ($N_{\rm hits}$) by neutrons, resulting in a product nucleus
\begin{equation} 
A^1 = A^0 - \sum_{j=0}^{N^0_{\rm hits}-1} \zeta^0(E^0,A^{j}) \ . 
\end{equation}
$E^0$ is the typical energy of neutrons impinging on the $k$=0 layer.
To produce a realistic distribution of product nuclei from this sub-layer, we draw $N^0_{\rm hits}$  and $\zeta^0$ from Poisson distributions peaking at $\bar{N^0}_{\rm hits}$ and $\bar{\zeta^0}$, where 
\begin{equation}
\bar{N^0}_{\rm hits}\sim \sigma_{\rm sp}\frac{N^{0}(1-e^{-1})}{4\pi R_{\rm in}^2}
\end{equation}
and $\bar{\zeta^0}$=$\bar{\zeta}(E^0,A^j)$. For subsequent sub-layers,
it follows  that
\begin{equation} 
\label{eq:k-layers} \bar{N}_{\rm hits}^k =  (1-e^{-1}) \bar{N}_{\rm hits}^{k-1} \bar{\zeta^k} \qquad {\rm and}\qquad E^k = \frac{(1-\eta) E^{k-1}}{\bar{\zeta^k} }  \  , 
\end{equation} 
where $\eta$ accounts roughly for the incident energy removed by radiation, nuclear excitation and other sub-products (pions, He, Deuterium etc ...). Since the pions and other multiplicities are small, we set $\eta=0$ \citep{Cugnon97}. 
 A Poissonian description is appropriate since we have a small spallation cross-section but a rapidly increasing $\bar{N}_{\rm hits}$ due
 to the cascading effect. The total number of layers that experience spallation
is given by $\min{(N_{\rm coll},k_{\rm max})}$ where $E^{k_{\rm max}-1}\simeq 73$ MeV, since for a specified thickness (mass) of target material, one can run out of material before spallation becomes insignificant. 

\section{Results}

{\it Spallation products in the inner SN ejecta}: Fig. 1 shows the probability distribution of the spallation products from $^{56}$Ni layer for $t_{\rm delay}$ of 4, 5 and 6 days. $^{56}$Ni is depleted, while $^{44}$Ti and light elements (H through Ne) are produced. The $t_{\rm delay}=5$ day Ni target case is particularly interesting for Cas A: note the Ti and C production peaks. Figure 2  shows normalized mass yields of spallated fragments $\eta_{\rm A}=M_{\rm A}/M_{\rm A_0}$, where $M_{\rm A_0}$ is the initial amount of Ni. We observe that for $3\ {\rm days} < t_{\rm delay} < 7$ days, the result is a {\it Ni-poor, Ti-rich, C-rich} debris. 

{\it Neutron-capture versus neutron-decay}: What happens to the neutrons themselves? For $E^{k_{\rm max}} < 73$ MeV following the spallation regime, the neutrons mainly deposit heat through inelastic collisions. Neutron-capture happens once the neutron energy is further reduced to $\le 30$ MeV. Compared to the free neutron lifetime of $\sim$ 720 seconds, the rather long n-capture timescale of $\tau_{\rm cap.}$=$1/(n_{\rm A}\sigma_{\rm cap.} v_{\rm n})\sim 10.2\ {\rm hours}$ (for fiducial values of $v_{\rm n}, M_A, \Delta R$ where $v_{\rm n}$ is the neutron thermal speed and $\sigma_{\rm cap.}$
the capture cross-section \citep{heil06}) implies that $t_{\rm delay}$ must be shorter than about 16 hours for significant neutron capture. In such cases, nucleosynthetic yields can be slightly altered from
those expected in a SN.

{\it Hydrogen Formation}: There are three sources of H in our model: (i) a direct byproduct of spallation (i.e. when $A$ is reduced to 1) for short time delays, which contributes at most 1\% of the total mass in the target layer; (ii) from spallation protons forming hydrogen via recombination once their energy $E\sim 73$ MeV (protons are about 1/3 of the total nucleons formed); (iii) from proton recombinations following $\beta$-decay of the neutrons that evade capture: for $t_{\rm delay} > 0.7$ days, most neutrons decay to protons which form H by recombination. The total amount of hydrogen that we estimate could form from these three factors is $M_{\rm H}\sim 0.1 M_{\odot} M_{\rm QN, -3}$.

\section{Conclusions and Predictions}
 
{\it Ni-poor, Ti-rich SNe}: We suggest that the Ti-rich Ni-poor yield of Cas A can be explained if the SN was followed by a QN with $t_{\rm delay} \sim 5$ days. From Fig.s 1\&2, this reproduces the observed abundance trends of key elements. In our model, $^{44}$Ti is formed from the destruction of $^{56}$Ni and thus Ti-rich dsQNe will be necessarily subluminous. Also, the rarity of Ti-rich SNe in the massive star populations could be because the mass-cut is above the Ti zone in these SNe, suppressing  Ti ejection.  {\it Only those experiencing a QN explosion following the SN, with the appropriate $t_{\rm delay}$, would show  $^{44}$Ti produced by spallation (as we suggest for Cas A)}. The higher mass-cut would also favor more massive NSs, ideal candidates for the QN transition to occur \citep{Staff06}. 
 
\begin{figure}[t!] \centering \includegraphics[width=0.5\textwidth]{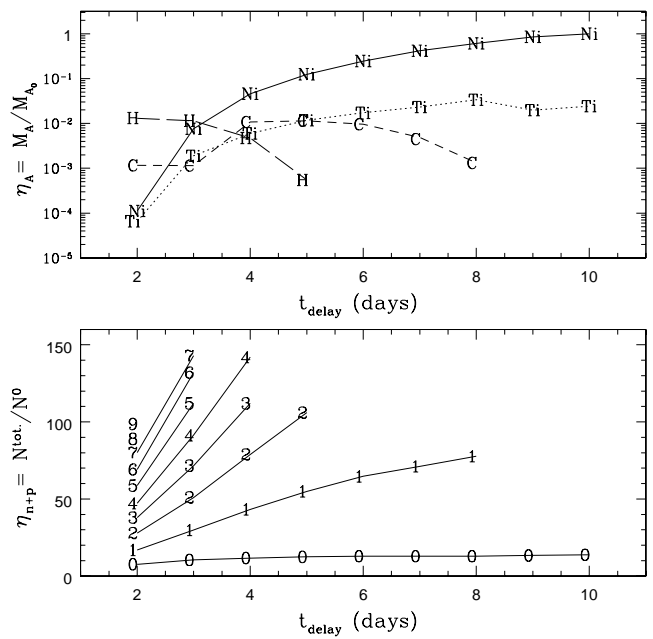} 

\caption{Mass Yields (upper panel) and total spallation neutrons\&protons (lower panel) versus time delay  for $^{56}$Ni target. In the lower panel, spallation layers are numbered 0-9, with spallation effectively ceasing after 10 layers for $t_{\rm delay}=2$ days.} \label{generations-Ni} \end{figure}

{\it Delayed Hydrogen signatures}: Decay of spallation neutrons to protons will not immediately form hydrogen since the recombination and thermal continuum  radiation remains trapped and ionizes the  H until it escapes when $\Delta R_{\rm A_0}= 1/(n_{\rm e, A_0}\sigma_{\rm Th.})$ where $\sigma_{\rm Th.}$ is the Thompson cross-section and $n_{\rm e, A_0}$ the electron density in the target layer. For $A_0=56$, the recombination radiation escapes when the innermost layers of the SN ejecta reaches a radius $R_{\rm in, esc.}\sim 1.8\times 10^{15}\ {\rm cm}\times M_{\rm A_0, 0.1}^{1/2}$. This implies that hydrogen should be observed $\sim 41\ {\rm days}\times M_{\rm A_0, 0.1}^{1/2}/v_{\rm in, 5000}$ days following the explosion.

{\it Other signatures}: The late time SN lightcurve should carry signatures of $^{56}$Ni and $^{56}$Co destruction by the QN ejecta. In the case of $^{56}$Ni  an amount of it would have decayed in $t_{\rm delay}$ and converted to $^{56}$Co, so that, for eg., only 57\% of the original $^{56}$Ni will experience spallation from the QN for $t_{\rm delay}=5$ days. The remaining  43\% of the original $^{56}$Ni would have decayed contributing mostly to PdV work in the SN ejecta. This means that the efficiency of the QN spallation determines how much of the remaining $^{56}$Ni is destroyed and directly influences the SN luminosity. In general, we find that  for $t_{\rm delay} < 8$ days, dsQNe appear as Ni-poor (sub-luminous) and Ti-rich/C-rich SNe. If $t_{\rm delay}\sim 8$ days, the Ni decay luminosity should be only slightly reduced but the Co decay luminosity will  be strongly suppressed (i.e. no 77 day tail). Even smaller $t_{\rm delay}$ can result in earlier destruction of $^{56}$Ni and thus lead to a subluminous SN with no 77 day Cobalt tail. It is interesting to consider the case of a Si layer (with $\sim 10$\% Ca; \citep{nomoto06}) over a Ni layer where the neutrons exit the Ni layer with $E_{\rm n} >> 73$ MeV (i.e. $k_{\rm max} >> N_{\rm coll}$). For fiducial values, this occurs for $t_{\rm delay} \sim 6$-8 days ($N_{\rm coll}\sim  2$; see Fig. 2, lower panel). This implies that the Si/Ca is destroyed by spallation, which  could explain the lack of lines such as Si, Ca and Iron in some superluminous SNe \citep{quimby11}. This delay is also ideal for explaining reenergization of superluminous SNe \citep{leahy08}. A $t_{\rm delay}\sim 5$ day dsQN leads to C formation (see Fig. 2); this may explain the C-rich atmosphere of the compact object in Cas A \citep{ho09}, which, in our model, would be a radio-quiet QS (an aligned rotator; \citep{ouyed06}) surrounded by a layer of fallback material rich in C. Finally, the QN-ejecta also contains heavy elements from neutron-capture (see \citep{jaikumar07}) and collisions between these and the inner layers of the SN ejecta will lead to other unique signatures which we have yet to investigate.

{\it Model assumptions}:  
Our model has some fine tuning that is unavoidable due to uncertainties regarding the nature of the hadron-quark phase transition. \citep{ho09} give mass-radius constraints for the compact object in Cas A, which effectively rule out low-mass QSs based on non-interacting quark equations of state. However, large and heavy QSs may exist, so long as the quark superconducting gap and strong coupling corrections are taken into account (e.g. \citep{alford07,kurkela2010,weissenborn11}). The issue of the mass-radius relation for quark stars is still a matter of debate. Spectral fitting of Cas A agrees very well with theoretical cooling models for NSs, when superfluidity and pair-breaking effects are taken into account \citep{shternin2011}. It is unlikely that a QS would exhibit exactly the same cooling behaviour as a NS, which is a problem for our model, but there is no comprehensive cooling simulation studies of QS and it might be purely coincidental that at this particular young age, a NS and a QS have the same surface temperature. Studies of cooling of QSs which include similar attention to physics details (e.g. color superconductivity) are needed to determine whether they could be at all consistent with Cas A.

{\it Acknowledgments:} This research  is supported by an operating grant from the
National Science and Engineering Research Council of Canada (NSERC).

\end{document}